\documentclass{ws-procs9x6}

\begin{document}

\title{Narrow Width of Penta-Quark Baryons \\
in QCD String Theory}

\author{Hideo~Suganuma and Hiroko~Ichie}

\address{
Tokyo Institute of Technology, Ohokayama, Meguro, Tokyo 152-8551, Japan\\
suganuma@th.phys.titech.ac.jp}

\author{Fumiko~Okiharu}

\address{
Department of Physics, Nihon University, Chiyoda, Tokyo 101-8308, Japan}

\author{Toru~T.~Takahashi}

\address{
YITP, Kyoto University, Kitashirakawa, Sakyo, Kyoto 606-8502, Japan}

\maketitle

\abstracts{
Using the QCD string theory, we investigate the physical reason of the 
narrow width of penta-quark baryons in terms of the large gluonic-excitation energy. 
In the QCD string theory, the penta-quark baryon decays via a gluonic-excited state  
of a baryon and meson system, where a pair of Y-shaped junction and anti-junction is created.
On the other hand, we find in lattice QCD that the lowest gluonic-excitation energy takes a 
large value of about 1 GeV.
Therefore, in the QCD string theory, the decay of the penta-quark 
baryon near the threshold is considered as a quantum tunneling process via a highly-excited state 
(a gluonic-excited state), which leads to an extremely narrow decay 
width of the penta-quark system.
}

\section{Introduction}

In 1969, Y.~Nambu first presented the string picture for hadrons\cite{N6970} 
to explain  the Veneziano amplitude\cite{V68} on the reactions and the resonances of hadrons.
Since then, the string picture has been one of the important pictures for hadrons 
and has provided many interesting ideas in the wide region of the particle physics.

Recently, various candidates of multi-quark hadrons 
have been experimentally observed.\cite{Theta,H1,NA49,X,Ds}
$\Theta^+$(1540),\cite{Theta} $\Xi^{--}(1862)$\cite{H1}
and $\Theta_c(3099)$\cite{NA49} are considered to be 
penta-quark (4Q-$\rm \bar Q$) states
and have been investigated with various theoretical 
frameworks.\cite{DPP97,OZ04,Z03,JW03,H03,SR03,KL03,STOI04,OST04,OST04p,BM04,NSTV04,BKST04,SZ04,KMN04}
X(3872)\cite{X} and $D_s(2317)$\cite{Ds} 
are expected to be tetra-quark (QQ-$\rm \bar Q \bar Q$) states\cite{CG03,CP04,S04}
from the consideration of their mass, narrow decay width and decay mode.

As a remarkable feature of multi-quark hadrons, their decay widths are extremely narrow, 
and it gives an interesting puzzle in the hadron physics.
In this paper, we investigate the physical reason of the narrow decay width of 
penta-quark baryons in the QCD string theory, with referring 
the recent lattice QCD results.\cite{STOI04,OST04,OST04p,TS01,TS02,TS03,TS04,STI04,IBSS03} 
With lattice QCD, we discuss the flux-tube picture and the gluonic excitation in Sects.2 and 3, respectively.
In Sect.4, we apply the QCD string theory to penta-quark dynamics, 
and try to estimate the decay width of the penta-quark baryon near the threshold.

\section{The Color-Flux-Tube Picture from Lattice QCD}

To begin with, we show the recent lattice QCD studies of the inter-quark potentials in 3Q, 4Q and 5Q 
systems,\cite{STOI04,OST04,OST04p,TS01,TS02} and 
revisit the color-flux-tube picture\cite{N74,KS75CKP83} for hadrons, 
which is idealized as the QCD string theory.

\subsection{The Three-Quark Potential in Lattice QCD}

For more than 300 different patterns of spatially-fixed 3Q systems, 
we calculate the 3Q potential from the 3Q Wilson loop 
in SU(3) lattice QCD with  
($\beta$=5.7, $12^3\times 24$),
($\beta$=5.8, $16^3\times 32$), 
($\beta$=6.0, $16^3\times 32$) and 
($\beta=6.2$, $24^4$).
For the accurate measurement, we construct the ground-state-dominant  
3Q operator using the smearing method.

To conclude, we find that the static ground-state 3Q potential $V_{\rm 3Q}^{\rm g.s.}$
is well described by the Coulomb plus Y-type linear potential, i.e., Y-Ansatz,  
\begin{eqnarray}
V_{\rm 3Q}^{\rm g.s.}=-A_{\rm 3Q}\sum_{i<j}\frac1{|{\bf r}_i-{\bf r}_j|}+
\sigma_{\rm 3Q}L_{\rm min}+C_{\rm 3Q},
\end{eqnarray}
within 1\%-level deviation.\cite{TS01,TS02,STI04}
Here,  $L_{\rm min}$ is the minimal value of the total length of the flux-tube, 
which is Y-shaped for the 3Q system.

To demonstrate this, we show in Fig.1(a) the 3Q confinement potential $V_{\rm 3Q}^{\rm conf}$, 
i.e., the 3Q potential subtracted by the Coulomb part, 
plotted against the Y-shaped flux-tube length $L_{\rm min}$.
For each $\beta$, clear linear correspondence is found between the 3Q confinement potential 
$V_{\rm 3Q}^{\rm conf}$ and $L_{\rm min}$, 
which indicates Y-Ansatz for the 3Q potential. 

\begin{figure}[h]
\begin{center}
\includegraphics[height=3.2cm]{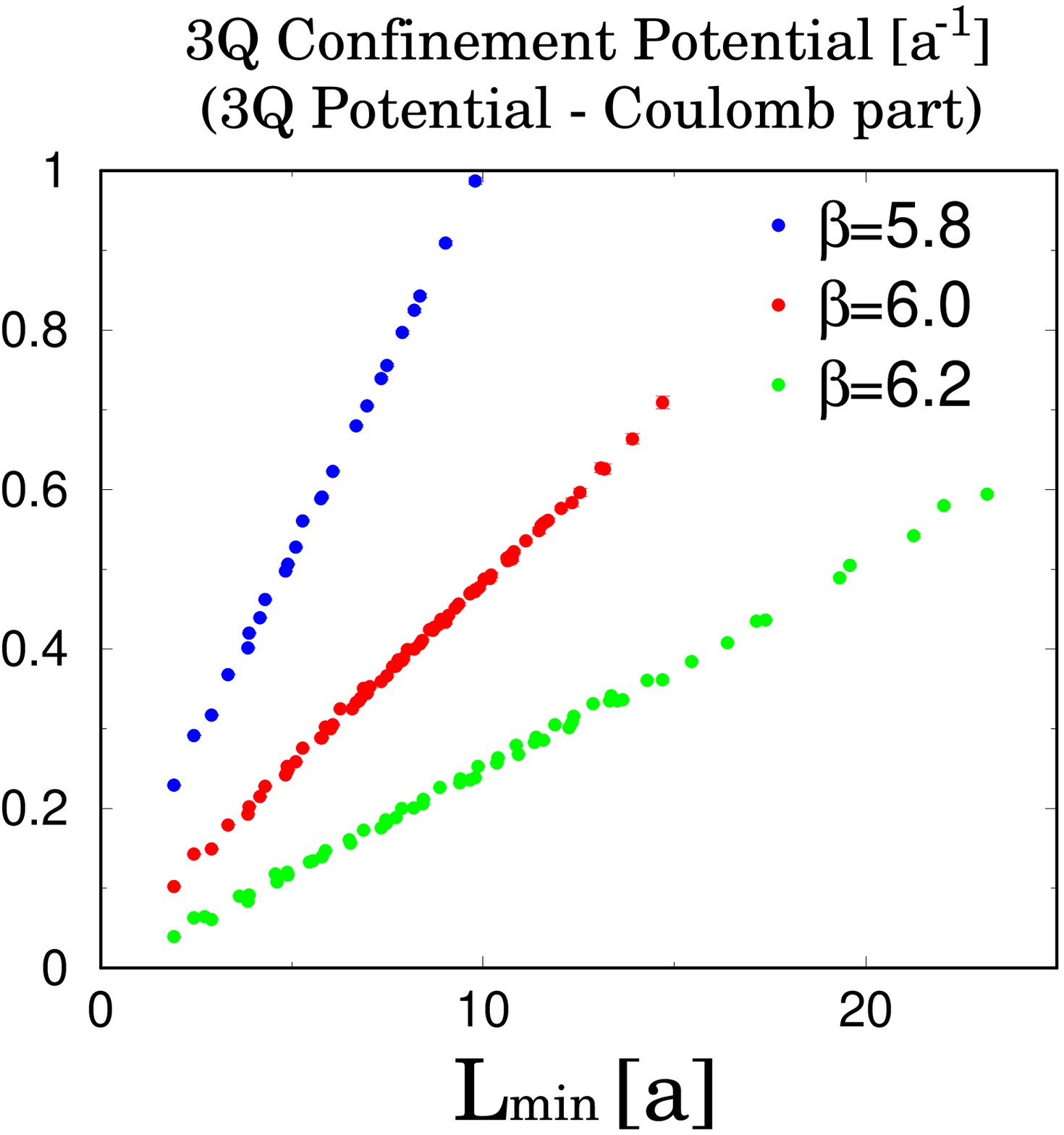}
\includegraphics[height=3.1cm]{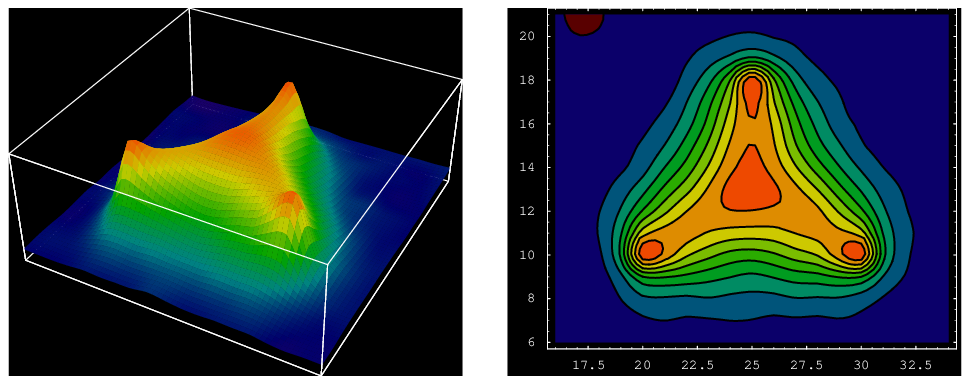}
\caption{
(a) The 3Q confinement potential $V_{\rm 3Q}^{\rm conf}$, 
i.e., the 3Q potential subtracted by the Coulomb part, 
plotted against 
the Y-shaped flux-tube length $L_{\rm min}$ 
in the lattice unit.
(b) The lattice QCD result for Y-type flux-tube formation 
in the spatially-fixed 3Q system.
The distance between the junction and each quark is about 0.5 fm.
}
\end{center}
\end{figure}

Recently, as a clear evidence for Y-Ansatz, 
Y-type flux-tube formation is actually observed 
in maximally-Abelian (MA) projected lattice QCD 
from the measurement of the action density  
in the spatially-fixed 3Q system.\cite{STI04,IBSS03}  

Thus, together with recent several other analytical and numerical studies,\cite{KS03,C04,BS04}
Y-Ansatz for the static 3Q potential seems to be almost settled. 
This result indicates the color-flux-tube picture for baryons.

\subsection{Tetra-quark and Penta-quark Potentials}

Motivated by recent experimental discoveries of multi-quark hadrons, 
we perform the first study of the multi-quark potentials in SU(3) lattice QCD. 
We calculate the multi-quark potentials from the multi-quark Wilson loops, 
and find that they can be expressed as 
the sum of OGE Coulomb potentials and the linear potential based on the flux-tube picture,\cite{STOI04,OST04,OST04p}
\begin{eqnarray}
V=\frac{g^2}{4\pi}\sum_{i<j}\frac{T^a_iT^a_j}{|{\bf r}_i-{\bf r}_j|}+\sigma L_{\rm min}+C,
\end{eqnarray}
where $L_{\rm min}$ is the minimal value of the total length of the flux-tube linking the static quarks.

Thus, the lattice QCD study indicates the color-flux-tube picture even for the multi-quark systems.
Also, this lattice result presents the proper Hamiltonian for the quark-model calculation of the multi-quark systems.

\section{The Gluonic Excitation in the 3Q System}

Next, we study the gluonic excitation in lattice QCD.\cite{TS03,TS04,STI04}
In the hadron physics, the gluonic excitation is one of the interesting phenomena 
beyond the quark model, and relates to the hybrid hadrons\cite{JKM0304,CP02} 
such as $q\bar qG$ and $qqqG$ in the valence picture. 
In QCD, the gluonic-excitation energy is given by the energy difference 
$\Delta E_{\rm 3Q} \equiv V_{\rm 3Q}^{\rm e.s.}-V_{\rm 3Q}^{\rm g.s.}$ 
between the ground-state potential $V_{\rm 3Q}^{\rm g.s.}$ and the excited-state potential $V_{\rm 3Q}^{\rm e.s.}$, 
and physically means the excitation energy of the gluon-field configuration in the static 3Q system.

For about 100 different patterns of 3Q systems, 
we perform the first study of the excited-state potential 
in SU(3) lattice QCD with $16^3\times 32$ at $\beta$=5.8 and 6.0 
by diagonalizing the QCD Hamiltonian in the presence of three quarks. 
In Fig.2, we show the 1st excited-state 3Q potential $V_{\rm 3Q}^{\rm e.s.}$ and 
the ground-state potential $V_{\rm 3Q}^{\rm g.s.}$.
The gluonic-excitation energy $\Delta E_{\rm 3Q} \equiv V_{\rm 3Q}^{\rm e.s.}-V_{\rm 3Q}^{\rm g.s.}$ 
in the 3Q system is found to be about 1GeV 
in the hadronic scale as $0.5{\rm fm} \le L_{\rm min} \le1.5{\rm fm}$.
This result predicts that the lowest hybrid baryon $qqqG$ has a large mass of about 2 GeV.

\begin{figure}[h]
\begin{center}
\includegraphics[height=4cm]{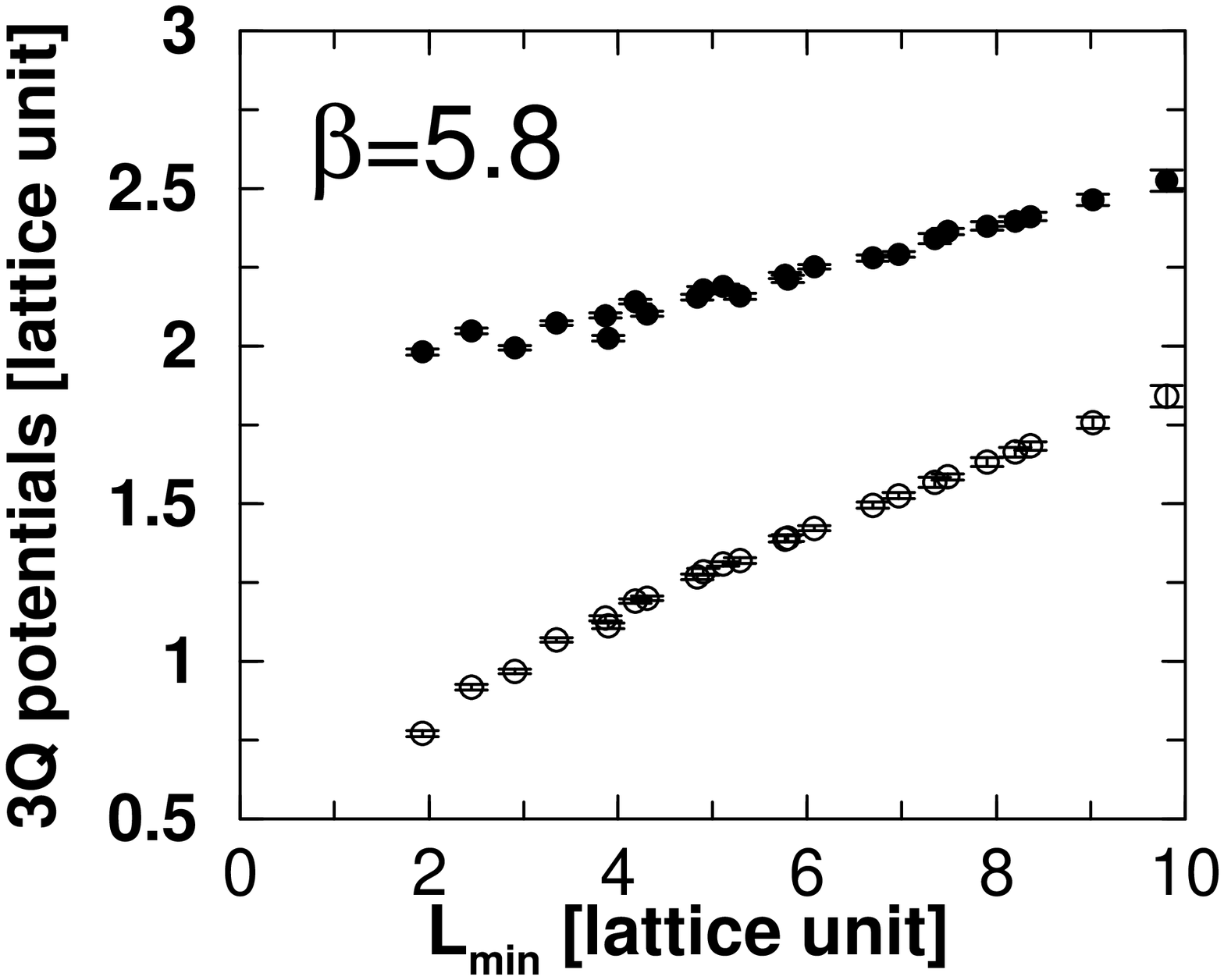}
\includegraphics[height=4cm]{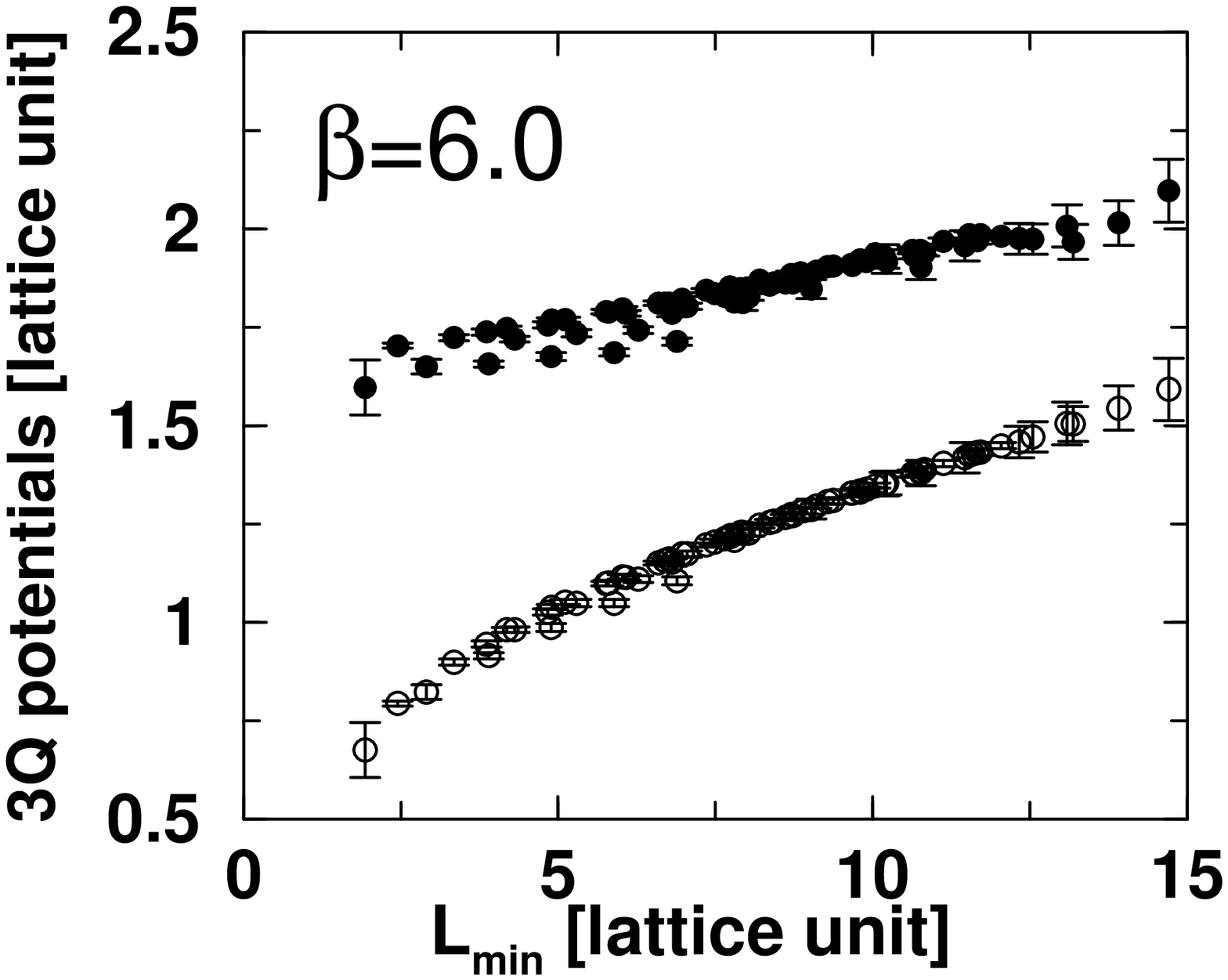}
\caption{
The 1st excited-state 3Q potential 
$V_{\rm 3Q}^{\rm e.s.}$ and 
the ground-state 3Q potential $V_{\rm 3Q}^{\rm g.s.}$.
The lattice results at $\beta=5.8$ and $\beta=6.0$ well coincide 
except for an irrelevant overall constant. 
The gluonic-excitation energy 
$\Delta E_{\rm 3Q} \equiv V_{\rm 3Q}^{\rm e.s.}-V_{\rm 3Q}^{\rm g.s.}$ 
is found to be about 1GeV in the hadronic scale 
as $0.5{\rm fm} \le L_{\rm min} \le 1.5{\rm fm}$.
}
\end{center}
\end{figure}

Note that the gluonic-excitation energy of about 1GeV is rather large compared with 
the excitation energies of the quark origin. 
Also for the $\rm Q \bar Q$ system, the gluonic-excitation energy is found to take a large value of about 1GeV\cite{JKM0304}.
Therefore, for low-lying hadrons, the contribution of gluonic excitations 
is considered to be negligible, and the dominant contribution is brought 
by quark dynamics such as the spin-orbit interaction, 
which results in the quark model without gluonic modes.\cite{STOI04,TS03,TS04,STI04} 
In Fig.3, we present a possible scenario from QCD to the massive quark model 
in terms of color confinement and dynamical chiral-symmetry breaking (DCSB).

\begin{figure}[h]
\begin{center}
\includegraphics[height=7.9cm]{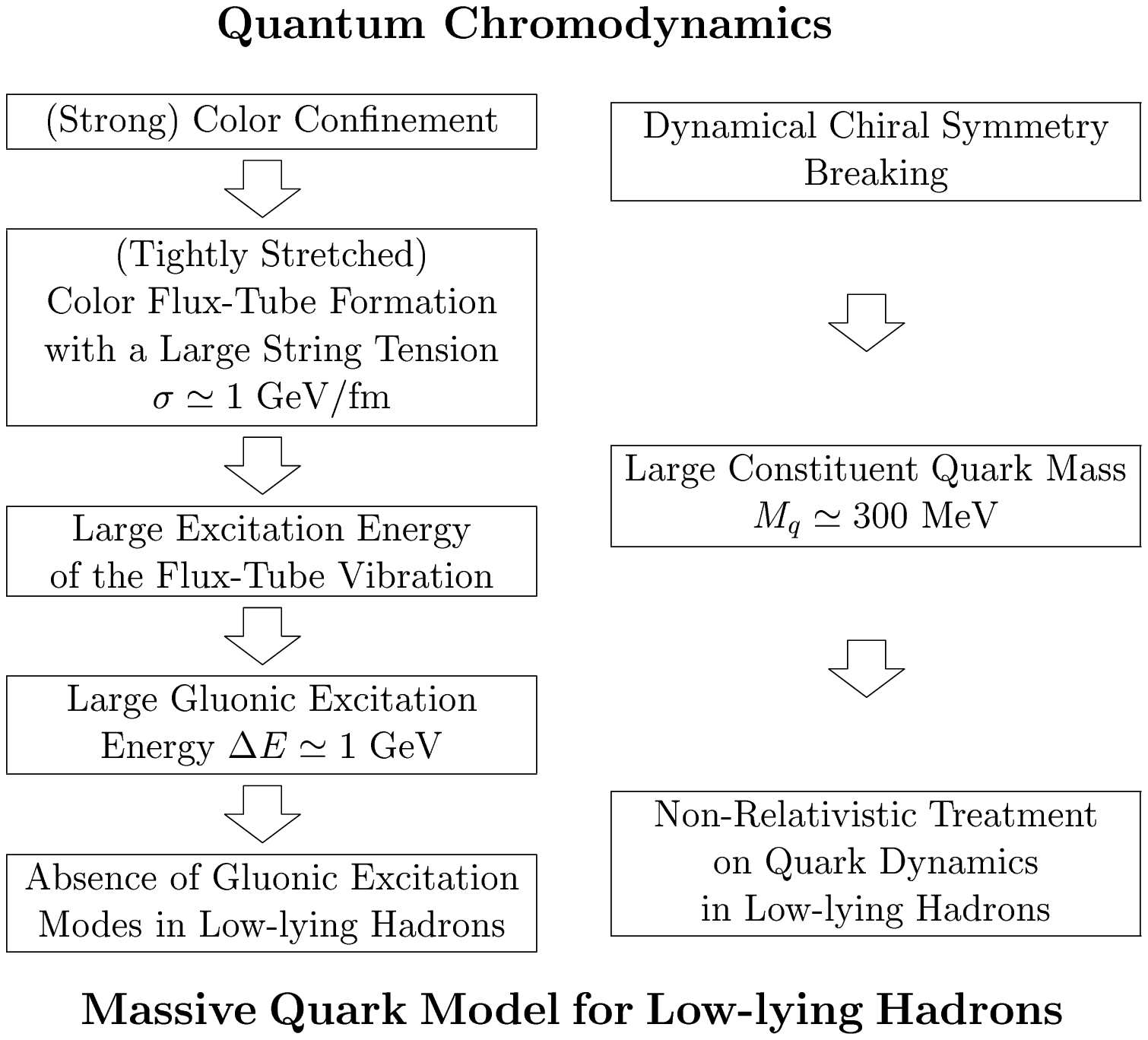}
\caption{A possible scenario from QCD to the quark model in terms of 
color confinement and DCSB.
DCSB leads to a large constituent quark mass of about 300 MeV, which enables the non-relativistic treatment 
for quark dynamics approximately. 
Color confinement results in the color flux-tube formation among quarks with a large string tension of $\sigma \simeq$ 1 GeV/fm.
In the flux-tube picture, the gluonic excitation is described as the flux-tube vibration, 
and its energy is expected to be large in the hadronic scale.
The large gluonic-excitation energy of about 1 GeV leads to 
the absence of the gluonic mode in low-lying hadrons, 
which plays the key role to the success of the quark model without gluonic-excitation modes.}
\end{center}
\end{figure}

\section{The QCD String Theory for the Penta-Quark Decay}

Our lattice QCD studies on the various inter-quark potentials 
indicate the flux-tube picture for hadrons, 
which is idealized as the QCD string model.
In this section, we consider penta-quark dynamics, 
especially for its extremely narrow width, in terms of the QCD string theory.

The ordinary string theory mainly describes open and closed strings corresponding to $\rm Q \bar Q$ mesons and glueballs, 
and has only two types of the reaction process as shown in Fig.4:
\begin{enumerate}
\item[1.] The string breaking (or fusion) process.
\item[2.] The string recombination process.
\end{enumerate}

\begin{figure}[h]
\begin{center}
\includegraphics[height=4cm]{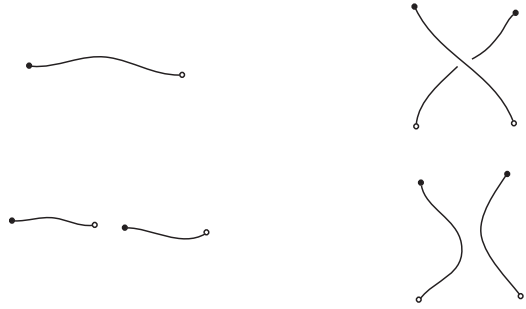}
\caption{The reaction process in the ordinary string theory:  
the string breaking (or fusion) process (left) and 
the string recombination process (right).}
\end{center}
\end{figure}

On the other hand, the QCD string theory describes also 
baryons and anti-baryons as the Y-shaped flux-tube, which  is 
different from the ordinary string theory.
Note that the appearance of the Y-type junction is peculiar to the QCD string theory with the SU(3) color structure.
Accordingly, the QCD string theory includes the third reaction process as shown in Fig.5:
\begin{enumerate}
\item[3.] The junction (J) and anti-junction ($\rm \bar J$) par creation (or annihilation) process.
\end{enumerate}
\begin{figure}[h]
\begin{center}
\includegraphics[width=10cm]{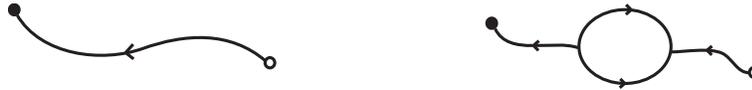}
\caption{
The junction (J) and anti-junction ($\rm \bar J$) par creation (or annihilation) process peculiar to the QCD string theory.}
\end{center}
\end{figure}
Through this J-$\bar {\rm J}$ pair creation process, 
the baryon and anti-baryon pair creation can be described.

As a remarkable fact in the QCD string theory, 
the decay/creation process of penta-quark baryons inevitably 
accompanies the J-$\bar {\rm J}$ creation\cite{BKST04} 
as shown in Fig.6.
Here, the intermediate state is considered as a gluonic-excited state, since it clearly corresponds to 
a non-quark-origin excitation. 

As shown in the previous section, 
the lattice QCD study indicates that 
such a gluonic-excited state is a highly-excited state with the excitation energy above 1GeV.
Then, in the QCD string theory, 
the decay process of the penta-quark baryon near the threshold 
can be regarded as a quantum tunneling, 
and therefore the penta-quark decay is expected to be strongly suppressed.
This leads to a very small decay width of penta-quark baryons.

\begin{figure}[h]
\begin{center}
\includegraphics[width=11.5cm]{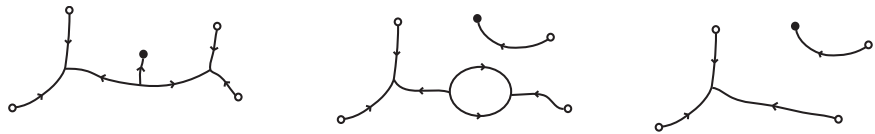}
\caption
{A decay process of the penta-quark baryon in the QCD string theory.
The penta-quark decay process inevitably accompanies the J-$\bar {\rm J}$ creation, which is a kind of the gluonic excitation.
There is also a decay process via the gluonic-excited meson.
}
\end{center}
\end{figure}

Now, we try to estimate the decay width of penta-quark baryons near the threshold 
in the QCD string theory. In the quantum tunneling as shown in Fig.6, 
the barrier height corresponds to the gluonic-excitation energy $\Delta E$ of the intermediate state, 
and can be estimated as $\Delta E \simeq$ 1GeV. 
The time scale $T$ for the tunneling process is expected to be the hadronic scale 
as $T =0.5 \sim 1{\rm fm}$, since $T$ cannot be smaller than the spatial size of the reaction area   
due to the causality. 
Then, the suppression factor for the penta-quark decay can be roughly estimated as 
\begin{eqnarray}
|\exp(-\Delta E T)|^2 \simeq |e^{-1{\rm GeV} \times (0.5 \sim 1){\rm fm}}|^2 \simeq 
10^{-2}\sim 10^{-4}.
\end{eqnarray}
Note that this suppression factor $|\exp(-\Delta E T)|^2$
appears in the decay process of low-lying penta-quarks 
for both positive- and negative-parity states. 

For the decay of $\Theta^+(1540)$ into N and K, the Q-value $Q$ is  
$Q= M(\Theta^+)-M({\rm N})-M({\rm K}) 
\simeq (1540-940-500) {\rm MeV} \simeq 100 {\rm MeV}$.
In ordinary sense, the decay width is expected to be controlled by  
$\Gamma_{\rm hadron} \simeq Q \simeq 100{\rm MeV}$.
Considering the extra suppression factor of $|\exp(-\Delta E T)|^2$, 
we get a rough order estimate for the decay width of $\Theta^+(1540)$ as 
\begin{eqnarray}
\Gamma[\Theta^+(1540)] \simeq \Gamma_{\rm hadron} \times |\exp(-\Delta E T)|^2 \simeq 1 \sim 10^{-2}{\rm MeV}.
\end{eqnarray}

\vspace{0.5cm}

\noindent
{\bf Acknowledgements}

\vspace{0.3cm}

\noindent
H.S. thanks Profs. T.~Kugo and A.~Sugamoto for useful discussions on the QCD string theory. 
H.S. is also grateful to Profs. K.~Hicks and C.~Hanhart for useful discussions on the penta-quark decay. 
The lattice QCD Monte Carlo simulations have been performed 
on NEC-SX5 at Osaka University.


\begin{thebibliography}{9}
\bibitem{N6970} Y.~Nambu, in {\it Symmetries and Quark Models} (Wayne State University, 1969); 
{\it Lecture Notes at the Copenhagen Symposium} (1970).
\bibitem{V68} G.~Veneziano,  {\it Nuovo Cim.} {\bf A57}, 190 (1968). 
\bibitem{Theta}
LEPS Collaboration (T.~Nakano {\it et al.}), 
{\it Phys. Rev. Lett.} {\bf 91}, 012002 (2003);
DIANA Collaboration (V.~V.~Narmin {\it et al.}), 
{\it Phys. Atom. Nucl.} {\bf 66}, 1715 (2003); 
CLAS Collaboration (S.~Stephanian {\it et al.}), 
{\it Phys. Rev. Lett.} {\bf 91}, 252001 (2003);
SAPHIR Collaboration (J.~Barth {\it et al.}), 
{\it Phys. Lett.} {\bf B572}, 127 (2003).
\bibitem{H1}
H1 Collaboration (A.~Aktas {\it et al.}), 
{\it Phys. Lett.} {\bf B588}, 17 (2004).
\bibitem{NA49}
NA49 Collaboration (C.~Alt {\it et al.}),
{\it Phys. Rev. Lett.} {\bf 92}, 042003 (2004).
\bibitem{X}
Belle Collaboration (S.~K.~Choi {\it et al.}),
{\it Phys. Rev. Lett.} {\bf 91}, 262001 (2003);
CDF II Collaboration (D.~Acosta {\it et al.}), 
{\it Phys. Rev. Lett.} {\bf 93}, 072001 (2004);
D0 Collaboration (V.~M.~Abazov {\it et al.}), {\it Phys. Rev. Lett.} {\bf 93}, 162002 (2004);
BABAR Collaboration (B.~Aubert {\it et al.}), {\it Phys. Rev. Lett.} {\bf 93}, 041801 (2004). 
\bibitem{Ds}
BABAR Collaboration (B.~Aubert {\it et al.}),
{\it Phys. Rev. Lett.} {\bf 90}, 242001 (2003);
Belle Collaboration (P.~Krokovny {\it et al.}),
{\it Phys. Rev. Lett.} {\bf 91}, 262002 (2003).
\bibitem{DPP97} D.~Diakonov, V.~Petrov and M.~Polyakov, 
{\it Z. Phys.} {\bf A359}, 305 (1997).
\bibitem{OZ04} For recent reviews, 
M.~Oka, {\it Prog. Theor. Phys.} {\bf 111}, 1 (2004); \\
S.~L.~Zhu, {\it Int. J. Mod. Phys.} {\bf A19}, 3439 (2004)
and their references. 
\bibitem{Z03} S.-L.~Zhu, {\it Phys. Rev. Lett.} {\bf 91}, 232002 (2003).
\bibitem{JW03} R.L.~Jaffe and F.~Wilczek, {\it Phys. Rev. Lett.} {\bf 91}, 232003 (2003).
\bibitem{H03} A. Hosaka, {\it Phys. Lett.} {\bf B571}, 55 (2003).
\bibitem{SR03} Fl. Stancu and D.O. Riska, {\it Phys. Lett.} {\bf B575}, 242 (2003). 
\bibitem{KL03} M.~Karliner and H.J.~Lipkin, hep-ph/0307243 (2003).
\bibitem{STOI04}
H.~Suganuma, T.~T.~Takahashi, F.~Okiharu and H.~Ichie, 
in {\it QCD Down Under}, March 2004, Adelaide, 
{\it Nucl. Phys.} {\bf B} (Proc. Suppl.) (2004) in press.
\bibitem{OST04}
F.~Okiharu, H.~Suganuma and T.~T.~Takahashi, 
``First study for the pentaquark potential in SU(3) lattice QCD", hep-lat/0407001.
\bibitem{OST04p}
F.~Okiharu, H.~Suganuma and T.~T.~Takahashi, this proceedings.
\bibitem{BM04} P.~Bicudo and G.M.~Marques, {\it Phys. Rev.} {\bf D69}, 011503 (2004); \\
P.~Bicudo, this proceedings.
\bibitem{NSTV04}
I.M.~Narodetskii, Yu. A.~Simonov, M.A.~Trusov and A.I.~Veselov, \\
{\it Phys. Lett.} B{\bf 578}, 318 (2004).
\bibitem{BKST04} M.~Bando, T.~Kugo, A.~Sugamoto and S.~Terunuma, this proceedings; \\
{\it Prog. Theor. Phys.} {\bf 112}, 325 (2004).
\bibitem{SZ04} X.-C. Song and S.-L. Zhu, {\it Mod. Phys. Lett.} {\bf A19}, 2791 (2004). 
\bibitem{KMN04} Y. Kanada-Enyo, O. Morimatsu and T. Nishikawa, this proceedings.
\bibitem{CG03}
F.~E.~Close and S.~Godfrey, {\it Phys. Lett.} {\bf B574}, 210 (2003). 
\bibitem{CP04} 
F.~E.~Close and P.~R.~Page, {\it Phys. Lett.} {\bf B578}, 119 (2004).
\bibitem{S04}
E.~S.~Swanson, {\it Phys. Lett.} {\bf B588}, 189 (2004); {\it ibid.} {\bf B598}, 197 (2004).
\bibitem{TS01} T.T.~Takahashi, H.~Matsufuru, Y.~Nemoto and H.~Suganuma, \\
{\it Phys. Rev. Lett.} {\bf 86}, 18 (2001);
in {\it Dynamics of Gauge Fields}, p.179 (2000).
\bibitem{TS02} T.T.~Takahashi, H.~Suganuma, Y.~Nemoto and H.~Matsufuru, \\
{\it Phys. Rev.} {\bf D65}, 114509 (2002); {\it AIP Conf. Proc.} {\bf 594}, 341 (2001).
\bibitem{TS03} T.T.~Takahashi and H.~Suganuma, {\it Phys. Rev. Lett.} {\bf 90}, 182001 (2003).
\bibitem{TS04} T.T.~Takahashi and H.~Suganuma, {\it Phys. Rev.} {\bf D70}, 074506 (2004).
\bibitem{STI04} H.~Suganuma, T.T.~Takahashi and H.~Ichie, 
in {\it Color Confinement and Hadrons in Quantum Chromodynamics}, p.249 
(World Scientific, 2004).
\bibitem{IBSS03} H.~Ichie, V.~Bornyakov, T.~Streuer and G.~Schierholz, \\
{\it Nucl. Phys.} {\bf A721}, 899 (2003); {\it Nucl. Phys.} {\bf B} (Proc. Suppl.) {\bf 119}, 751 (2003). 
\bibitem{N74} Y.~Nambu, {\it Phys. Rev.} {\bf D10}, 4262 (1974).
\bibitem{KS75CKP83} J.~Kogut and L.~Susskind, {\it Phys. Rev.} {\bf D11}, 395 (1975);
J.~Carlson, J.~Kogut and V.~Pandharipande, {\it Phys. Rev.} {\bf D27}, 233 (1983); {\it ibid.} {\bf D28}, 2807 (1983).
\bibitem{KS03} D.S. Kuzmenko and Yu.~A. Simonov, {\it Phys. Atom. Nucl.} {\bf 66}, 950 (2003).
\bibitem{C04} J.M.~Cornwall, {\it Phys. Rev.} {\bf D69}, 065013 (2004).
\bibitem{BS04} P.O.~Bowman and A.P.~Szczepaniak, {\it Phys. Rev.} {\bf D70}, 016002 (2004).
\bibitem{JKM0304}
K.J.~Juge, J.~Kuti and C.~Morningstar, {\it Phys. Rev. Lett.} {\bf 90}, 161601 (2003); 
{\it AIP Conf. Proc.} {\bf 688}, 193 (2004).
\bibitem{CP02}
S.~Capstick and P.R.~Page, {\it Phys. Rev.} {\bf C66}, 065204 (2002). 
\end{thebibliography}
\end{document}